# Understanding effects of TCO work function on the performance of organic solar cells by numerical simulation

Aqing Chen [1*], Kaigui Zhu [2], Qingyi Shao [3]

[1] *College of Materials & Environmental Engineering, Hangzhou Dianzi University, Hangzhou 310018, P R China.*

[2] *Department of physics, Beihang University, Beijing 100191, P R China.*

[3] *School of Telecommunication Engineering, South China Normal University, Guangzhou 510006, China*

**Abstract**

The influences of work function of transparent conducting oxides (TCO) on the performance of organic solar cells, including open circuit voltage, conversion efficiency and fill factor, has been simulated. It is obtained that for non-Ohmic contact the open circuit voltage and conversion efficiency increase monotonically with the TCO work function but keep constant for Ohmic contact. Fill factor decreases and increases with the electrode work function when the electrode work function is below and above a critical value (4.2 eV for TCO and 4.5 eV for back-contact), respectively. The results of this simulation are significant in the choice of TCO contacts to optimize organic planar heterojunction solar cells.

[*] Corresponding author.
E-mail address: aqchen@hdu.edu.cn

# 1. Introduction

Organic solar cells (OSCs) compared to the silicon based solar cells have gained intensive attraction in the past two decades due to their advantages of light-weight, low cost and easy-making [1–6]. Understanding the device physics and optimizing the device parameters are of an utmost importance to enhance the performance of OSCs. The electrode work function is considered to be a critical factor of the performance of OSCs. Numerous investigations on the effects of cathode work function on the performance of OSCs have been done [4,7–9]. It was reported that the open-circuit voltage ($V_{OC}$) was independent much on the cathode work functions [4], varying very small (<0.2 V) in the wide work function range of 2.87 eV (Ca) to 4.28 eV (Au)[8], but the cell resistance depends strongly on it [10]. As another key device parameter, the work function of transparent conducting oxides (TCO) also significantly impacts the performance of OSCs [11] because the TCO work function influences the band alignment between TCO and organic active layer. Scientists modified TCO layer to improve the power conversion efficiency of OSCs. For instance, Lee et al. produced $MoO_3$ graded ITO anodes to enhance the power conversion efficiency by 3.6 % [12]. Lei et al. enhanced the work function of ITO anodes with CuS to achieve a high power conversion efficiency of 7.4 % [13]. However, there is few systematical investigations to give insight to the effects of TCO work function on the performance of OSCs, as far as we know. Understanding the physical mechanisms on the influence of TCO work function on the performance of OSCs is important for the enhancement of the performance.

In this present work, the relationship between TCO work function and the performance of OSCs, including conversion efficiency (η), open circuit voltage ($V_{OC}$), and fill factor (FF), is investigated in depth by numerical simulation which is a method for the understanding and design of solar cells [14,15]. It is shown that both the $V_{OC}$ and η increase monotonically with the TCO work function if the non-ohmic contact is formed, and keep constant once ohmic contact is achieved. The FF has a complex relationship with the TCO work function depending on the contribution of contact potential and the carrier concentration. The hole concentration, band diagram and quantum efficiency were calculated for further explanation on the results.

## 2. Simulation structure of organic solar cells and calculation methods

Figure 1 shows the simulation model of organic solar cells. The TCO window layers can be ZnO, ITO, $SnO_2$ or the layer including ITO and PEDOT:PSS. The active layer consists of the Poly (3-hexylthiophene) (P3HT) and [6, 6]-phenyl C61 butyric acid methylester (PCBM). Thickness of both P3HT and PCBM is set to 100 nm. The generation layer between P3HT and PCBM is simulated with a 1 nm P3HT:PCBM (1:1) blend. In order to study the effects of TCO work function on the OSCs, the wxAMPS software [16] based on the AMPS-1D program which is a general solar cell simulation program [17] was employed to simulate the organic cells. In AMPS-1D there is no consideration of the interface recombination, but in wxAMPS two tunneling models are incorporated to consider tunneling effects, namely, the tap-assisted tunneling [18] and the intra-band tunneling [19]. In tap-assisted tunneling model, the recombination

rate is written as [18]

$$R_{trap} = N_T \frac{c_n c_p n_t p_t - e_n e_p}{c_n n_t + c_p p_t + e_n + e_p} \quad (1)$$

where $N_T$ is the trap density. $n_t$ and $p_t$ are the densities of free electron and hole at the trap location. $c_n$ and $e_n$ are the electron capture and emission rates. $c_p$ and $e_p$ are the hole capture and emission rates. The capture rates $c_{n,p}$ indicate the capture of an electron or hole at the trap location and the emission rates $e_{p,p}$ give the probabilities per unit time for a captured carrier to escape from the trap. when an weak electric field is present, the carrier densities at a certain location are obtained by conventional density, but when the electric field is strong, the density of carriers at tap location is increased resulting from the finite probability of carriers tunneling into the gap. In this simulation, the trap-assisted tunneling model which is necessary for the junction with high electric field [16,18] is chose to obtain more precise simulation results because high electric field at our case will be obtained when TCO work function is changed from a small value to a high value. The electron and hole mobility of P3HT:PCBM blend films, as well as the effective band gap $E_g$, inputted into wxAMPS are obtained from the ref [20], as listed in Table I. TCO like $In_2O_3$, $SnO_2$, ZnO, and $Cu_2O$ have different work functions due to the concentration of crystallographic defects like oxygen vacancies [21]. For example, ITO films deposited by RF sputtering have a varied work function from 3.3 eV and 5.7 eV depending on the conditions of the deposition process [22–24]. So, it is reasonable to set the TCO work function range of 3.8 to 5.2 eV in this simulation. The band alignment is schematically described in Fig. 1 (a). The absorption of active layer which determines the electron and hole generation, it is, hence, important to input the experimental absorption coefficient of P3HT:PCBM blend [25] into the wxAMPS. Some key parameters of P3HT, P3HT:PCBM and PCBM used for this simulation are list in the following table I. Because this work only focuses on the effects of TCO on the performance of organic solar cells, the reflectivity of window layer and rear contact are assumed ideally as zero and one, respectively. Both the surface recombination velocities of electrons and holes were set as 1.0 $\times 10^7$ cm/s. The AM 1.5 spectrum normalized to 100 mW/cm$^2$ is used as this simulation illumination condition.

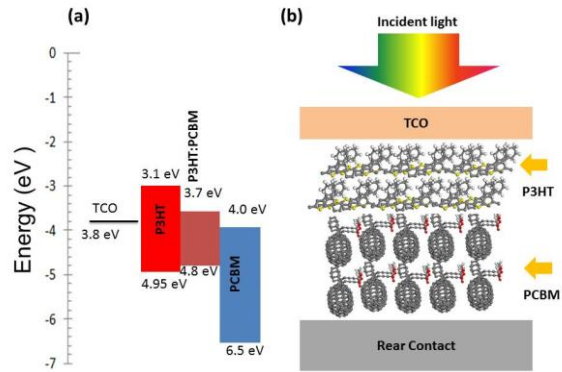

Fig.1 (a) the band alignment and (b) the simulation mode of organic solar cells.

Table 1 Some key parameters of P3HT, P3HT:PCBM and PCBM

| Parameter | P3HT | P3HT:PCBM | PCBM |
| --- | --- | --- | --- |
| Permittivity | 3.0 | 3.0 | 3.0 |
| Thickness (nm) | 100 | 1 | 100 |
| Effective Conduction Band Density (cm$^{-3}$) | $1 \times 10^{22}$ | $2.5 \times 10^{20}$ | $1 \times 10^{22}$ |
| Effective Valence Band Density (cm$^{-3}$) | $1 \times 10^{22}$ | $2.5 \times 10^{20}$ | $1 \times 10^{22}$ |
| $\mu_n$ (cm$^2$/Vs) | $1 \times 10^{-4}$ | $8.78 \times 10^{-4}$ | $1 \times 10^{-3}$ |
| $\mu_p$ (cm$^2$/Vs) | $1 \times 10^{-3}$ | $3.47 \times 10^{-4}$ | $1 \times 10^{-4}$ |

| Electron affinity (eV) | 3.1 | 3.7 | 3.7 |
| --- | --- | --- | --- |
| Band gap (eV) | 1.85 | 1.1 | 2.1 |
| Bulk defect properties | Donor-like gap states at $E_V = 0.93$ eV $N_{TD}=1\times10^{10}$ cm$^{-3}$ $n=1\times10^{-9}$ cm$^2$ $p=1\times10^{-10}$ cm$^2$ Acceptor-like gap states at $E_V= 0.93$ eV $N_{TA}=1\times10^{10}$ cm$^3$ $n=1\times10^{-10}$ cm$^2$ $p=1\times10^{-9}$ cm$^2$ | | Donor-like gap states at $E_V = 1.05$ eV $N_{TD}=1\times10^{10}$ cm$^{-3}$ $n=1\times10^{-9}$ cm$^2$ $p=1\times10^{-15}$ cm$^2$ Acceptor-like gap states at $E_V= 1.05$ eV $N_{TA}=1\times10^{10}$ cm$^3$ $n=1\times10^{-10}$ cm$^2$ $p=1\times10^{-9}$ cm$^2$ |

First, in order to explore the effects of the TCO (front-contact) work function ($W_{TCO}$) on the performance of organic solar cells, the difference between the work function of back-contact and PCBM is kept a constant value of 0.625 eV so that there is no built-in potential between back-contact and PCBM at thermodynamic equilibrium [9]. The open circuit voltage of the organic solar cells is influenced by many factors including the temperature, light density and work function of the electrode [26]. With respect to our case, we only consider the effects of the work function of the electrodes on the open circuit voltage at temperature of 300 K and under one AM 1.5 spectrum.

3. Analyses and results

As can be seen in Fig. 2 (a), the open circuit voltage increases linearly with the TCO work function when $W_{TCO} < 4.725$ eV. For a low TCO work function, contact potential between TCO and P3HT becomes high, so the hole should cross a barrier to arrive at the TCO. When the barrier is too high (> 0.25 eV), the non-Ohmic contact between TCO and P3HT will be formed, resulting in non-zero interfacial electric field at the hole injecting metal/organic interface. Consequently, the open circuit voltage is mainly determined by the TCO work function and proportion to the TCO work function for non-ohmic contact, as shown in Fig.2 (a). The band diagram of this device at different TCO work function (when $W_{TCO} < 4.725$ eV) can illustrate this relationship clearly, as shown in Fig.3 (a). It is obvious to see that the bands bend upward as the $W_{TCO}$ increases. But when the TCO work function reaches a high value of 4.75 eV, the difference between TCO work function and the HOMO level of P3HT (4.95 eV) is less than 0.2 eV. The hole, hence, can transfer into the P3HT layer by thermal activation. Therefore, the contact between TCO and P3HT turns to be Ohmic contact. Under this condition, the Fermi energy level of TCO can be pinned to the HOMO level of the P3HT (donor) [27]. This can be demonstrate by the band diagram when $W_{TCO} > 4.625$ eV, as sown in fig. 3 (b). The band is independent on the TCO work function. Therefore, the open circuit voltage of this organic solar cell is governed by the difference between the HOMO level of donor and LUMO level of accept and independent on the work function of electrodes [8]. But the open circuit voltage is less than the difference between HOMO level of donor (4.85 eV) and LUMO level of acceptor (3.7 eV) due to the carrier recombination and the disorder in organic solar cells, which can be described by the following equation,

$$qV_{OC} = \Delta E_{DA} - \frac{\sigma^2}{k_BT} - k_BTln\left(\frac{N_AN_D}{np}\right) \qquad (2)$$

where $\Delta E_{DA}$ is the difference between HOMO of donor and LUMO of acceptor, the second and third term are the contribution of disorder and carrier recombination, respectively. As shown in Fig. 2 (a), the open circuit voltage varies by about 0.02 V in the TCO work function range of 4.7 eV to 5.2 eV and is approximate 1.02 V which is close to the experimental value (~ 0.9 V) of organic cells with the front electrodes of ITO [28]. It also can be seen in Fig. 2 that the conversion efficiency has the same characteristics as open circuit voltage. The carrier concentration in our case is set to a small value on the order of $10^{11}$ cm$^{-3}$. That means that the serial resistance in the present simulated cells is large. The fill factor, hence, becomes small due to the large serial resistance [10,29], as shown in fig. 2 (b). But, it is worth noting that the fill factor decreases with TCO work function in the range of 3.8 eV to 4.2 eV, then increases with the TCO work function in the range of 4.2 eV to 4.8 eV.

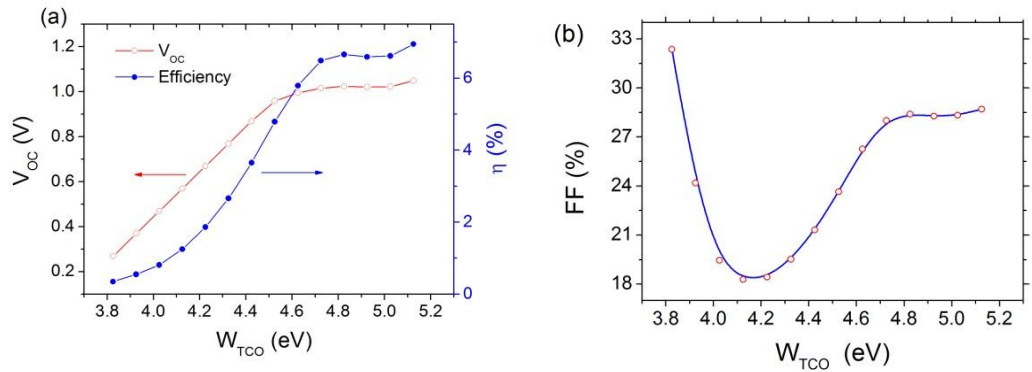

Fig. 2 (a) $V_{OC}$ and $\eta$ as function of TCO work function, (b) FF as a function of TCO work function.

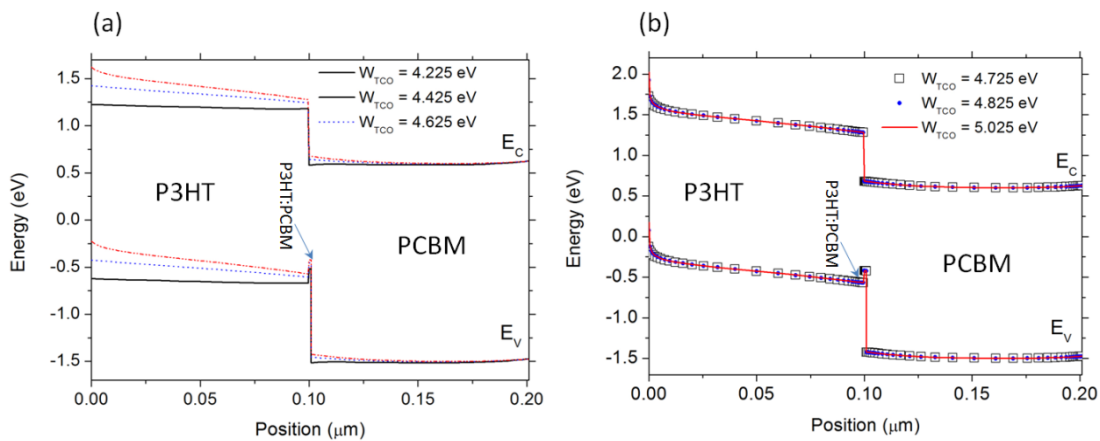
8

Fig. 3 band diagram of the OSCs at different TCO work function. (a) is the band diagram when $W_{TCO} < 4.625$ eV and (b) is the band diagram when $W_{TCO} > 4.625$ eV.

In order to explore the novel relationship between FF and $W_{TCO}$, the hole concentration as a function of the TCO work function were calculated. As can be seen in Fig. 4 (a), the hole concentration is nearly unchanged at the interface with a low value of $1.0 \times 10^{11}$ cm$^{-3}$ when the TCO work function varies from 3.825 eV to 4.2 eV. That means that contact resistance is independent on the hole concentration in the TCO work function range of 3.825 eV to 4.2 eV. As discussed above, the contact between TCO and P3HT is non-Ohmic contact when TCO work function is less than 4.2 eV. Therefore, it can be deduced that the contact resistance is dominated by the work function difference between front and back electrode. Increasing TCO work function, hence, increases the contact resistance. The large contact resistance increases the series resistance of this organic solar cell. But when $W_{TCO} > 4.2$ eV, the hole concentration at the interface between TCO and P3HT increases sharply from $3 \times 10^{11}$ cm$^{-3}$ to $1 \times 10^{15}$ cm$^{-3}$, as shown in Fig. 4 (b). The high hole concentration can result in tunneling easily. Simultaneously, the contact potential at the interface between TCO and P3HT decreases as the TCO work function increases. Therefore, the decreases of contact barrier and the increases of free hole concentration ultimately lead to a low series resistance due to the increase of TCO work functions. The influence of series resistance $R_S$ on the performance of OSCs is to reduce the fill factor *FF*. As can be seen in Fig. 2 (b) and discussed above, the TCO work function of 4.2 eV is the turning point regards to the *FF*. As been pointed out above, the contact between TCO and P3HT is Ohmic contact for $W_{TCO} > 4.725$. That means the contribution of the contact barrier to the series resistance can be neglected in the work function range of 4.725 eV to 5.2 eV. So, the FF is nearly unchanged due to the constant series resistance which is mainly determined by the resistance of P3HT and PCBM when the TCO work function varies from 4.725 eV to 5.2 eV.

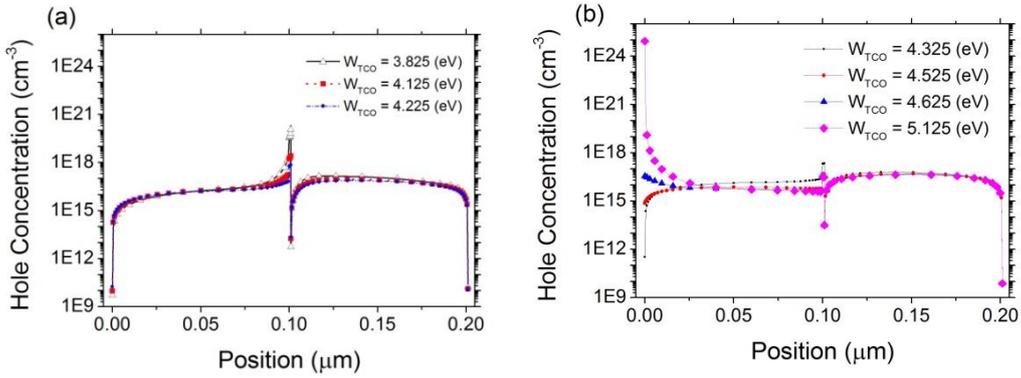

Fig. 4 the hole concentration at the different TCO work function (a) below 4.225 eV (b) above 4.325 eV.

Furthermore, the quantum efficiency QE, the ratio of the number of electrons in the external circuit produced by an incident photon of a given wavelength [30] was calculated to give an insight into the effects of TCO work function on the performance of this organic solar cells. Fig. 5 shows the quantum efficiency in the wavelength range of 350 nm to 650 nm at various TCO work function, which agrees well with the experimental results [5]. The photo-generated current $I_{ph}$ strongly depends on the quantum efficiency, as described by the following equation.

$$I_{ph} = q \int_{(\lambda)} \phi(\lambda)\{1 - R(\lambda)\} QE(\lambda) d\lambda \qquad (3)$$

where $\phi(\lambda)$ is the photo flux incident on the cell at wavelength and $QE(\lambda)$ is the quantum efficiency, $R(\lambda)$ is the reflectance on the front surface. In Fig.5, it is obvious observed that the quantum efficient is enhanced rapidly when the TCO work function increases from 3.825 eV to 4.525 eV, and then increases slowly at a high TCO work function. Therefore, high TCO work function leads to larger photo-generated current according to the equation 3. The analyses on quantum efficiency give a good interpretation on the variation of the $V_{OC}$ and η as a function of TCO work function.

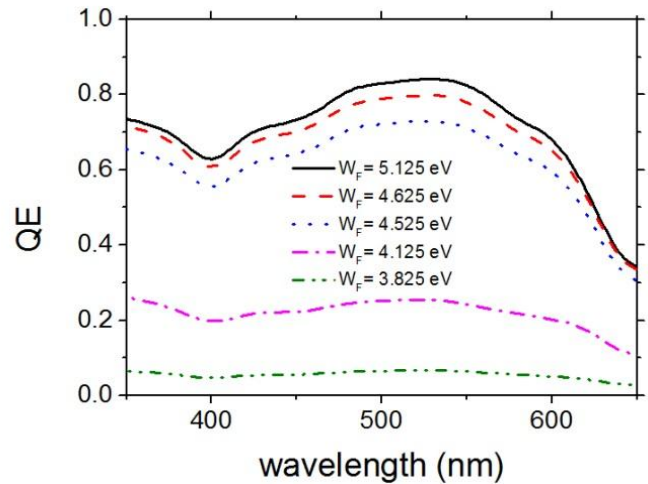

Fig. 5 quantum efficiency at different TCO work function.

## 4. Conclusion

In summary, the effects of TCO work function on the performance of organic solar cells have been studied. When the non-Ohmic contact is formed, the open circuit voltage and conversion efficiency increase monotonically with the TCO work function but keep constant once Ohmic contact is achieved. It is novel to find that the fill factor does not vary monotonically with TCO work function. Fill factor decreases with the $W_{TCO}$ for $W_{TCO} < 4.2$ eV and then increases with the $W_{TCO}$ for $W_{TCO} > 4.2$ eV. The results are suggested that increasing the TCO work function is a useful way to improve the performance of organic solar cells. Therefore, this work has a significant reference of choosing proper TCO to fabricate the organic solar cells with a high performance.

**Acknowledgement**

The authors greatly acknowledge Professer S. Fonash of the Pennsylvania State University and the Electric Power Research Institute for providing the AMPS-1D program, and Yiming Liu of Nankai University for providing the wxAMPS software used in this simulation.